\documentclass[aps,prb,superscriptaddress,twocolumn,showpacs]{revtex4}

\usepackage{graphicx}
\usepackage{dcolumn}
\usepackage{bm}

\begin{document}

\title{Low-temperature heat transport and magnetic-structure transition
of the hexagonal TmMnO$_3$ single crystals}

\author{X. M. Wang}
\affiliation{Hefei National Laboratory for Physical Sciences at
Microscale, University of Science and Technology of China, Hefei,
Anhui 230026, People's Republic of China}

\author{Z. Y. Zhao}
\affiliation{Hefei National Laboratory for Physical Sciences at
Microscale, University of Science and Technology of China, Hefei,
Anhui 230026, People's Republic of China}

\author{C. Fan}
\affiliation{Hefei National Laboratory for Physical Sciences at
Microscale, University of Science and Technology of China, Hefei,
Anhui 230026, People's Republic of China}

\author{X. G. Liu}
\affiliation{Hefei National Laboratory for Physical Sciences at
Microscale, University of Science and Technology of China, Hefei,
Anhui 230026, People's Republic of China}

\author{Q. J. Li}
\affiliation{Hefei National Laboratory for Physical Sciences at
Microscale, University of Science and Technology of China, Hefei,
Anhui 230026, People's Republic of China}

\author{F. B. Zhang}
\affiliation{Hefei National Laboratory for Physical Sciences at
Microscale, University of Science and Technology of China, Hefei,
Anhui 230026, People's Republic of China}

\author{L. M. Chen}
\affiliation{Department of Physics, University of Science and
Technology of China, Hefei, Anhui 230026, People's Republic of
China}

\author{X. Zhao}
\affiliation{School of Physical Sciences, University of Science
and Technology of China, Hefei, Anhui 230026, People's Republic of
China}

\author{X. F. Sun}
\email{xfsun@ustc.edu.cn} \affiliation{Hefei National Laboratory
for Physical Sciences at Microscale, University of Science and
Technology of China, Hefei, Anhui 230026, People's Republic of
China}

\date{\today}

\begin{abstract}

We study the low-temperature heat transport, as well as the
magnetization and the specific heat, of TmMnO$_3$ single crystals
to probe the transitions of magnetic structure induced by magnetic
field. It is found that the low-$T$ thermal conductivity
($\kappa$) shows strong magnetic-field dependence and the overall
behaviors can be understood in the scenario of magnetic scattering
on phonons. In addition, a strong ``dip"-like feature shows up in
$\kappa(H)$ isotherms at 3.5--4 T for $H \parallel c$, which is
related to a known spin re-orientation of Mn$^{3+}$ moments. The
absence of this phenomenon for $H \parallel a$ indicates that the
magnetic-structure transition of TmMnO$_3$ cannot be driven by the
in-plane field. In comparison, the magnetothermal conductivity of
TmMnO$_3$ is much larger than that of YMnO$_3$ but smaller than
that of HoMnO$_3$, indicating that the magnetisms of rare-earth
ions are playing the key role in the spin-phonon coupling of the
hexagonal manganites.

\end{abstract}


\maketitle

\section{INTRODUCTION}

The hexagonal manganites $R$MnO$_3$ ($R$ = Y, Ho, Er, Tm, Yb, and
Lu) have been extensively studied because of their
multiferroicity.\cite{Fiebig1, Lottermoser, Aken, Fiebig2,
Fiebig3, Yen1, Yen2, Vajk, Skumryev, Hur, Ueland, Standard} In
these materials, the ferroelectric order is formed at temperatures
as high as 800 K, which results from the ionic displacements
breaking the inversion symmetry of the lattice. The Mn$^{3+}$
moments develop the antiferromagnetic (AF) ordering below
$T_{N,Mn}$ = 70--90 K. The magnetoelectric coupling has been found
to be strong between the $c$-axis polarization and the $ab$-plane
staggered AF magnetization at low temperatures, as the strong
dielectric anomalies at the magnetic phase transitions have
demonstrated.\cite{Iwata, Yen1, Hur} Naturally, the magnetic
structures and their transitions driven by changing temperature or
magnetic field\cite{Fiebig3, Yen2} are the central issues of the
magnetoelectric coupling in these compounds, which are still not
fully understood due to the complexity of Mn$^{3+}$ and $R^{3+}$
magnetism.

The Mn$^{3+}$ ions form triangular planar sublattices and the AF
exchange coupling among Mn$^{3+}$ moments is therefore
geometrically frustrated. As a result, below $T_{N,Mn}$, the
Mn$^{3+}$ moments are ordered in a configuration that the
neighboring moments are 120$^{\circ}$ rotated, with the space
group $P6'_3c'm$.\cite{Koehler} However, the homometric
configurations of the Mn$^{3+}$ moments in the triangular lattice
are possible to change from each other when lowering temperature
or applying magnetic field.\cite{Fiebig3, Yen1, Yen2, Hur, Lorenz,
Brown} The situation would be more complicated for those
$R$MnO$_3$ with their rare-earth ions having magnetic moments.
ErMnO$_3$, YbMnO$_3$, and HoMnO$_3$ were found to have a second AF
transition at 2.5--4.6 K, below which the long-range AF order of
$R^{3+}$ ions are formed.\cite{Yen2} This low-$T$ AF state can be
easily destroyed by a weak magnetic field for ErMnO$_3$ and
YbMnO$_3$, while it is much more stable in magnetic field for
HoMnO$_3$.\cite{Yen2} In contrast, TmMnO$_3$ does not display the
long-range order of Tm$^{3+}$ ions although they have pretty large
moments.\cite{Yen2, Skumryev, Salama} Furthermore, the $R^{3+}$
moments were believed to orientate along the $c$ axis with an
Ising-like anisotropy but their spin structures are not simple
because the $R^{3+}$ ions present in two different
crystallographic sites, 2$a$ and 4$b$.\cite{Yen2, Lonkai, Nandi,
Fabreges1, Salama} Because of the magnetic interaction between the
rare-earth ions and the Mn$^{3+}$ ions, the field-induced
transitions of Mn$^{3+}$ magnetic structure and the resulted $H-T$
phase diagrams are strongly dependent on the magnetisms of
rare-earth ions.\cite{Fiebig3, Yen1, Yen2, Hur, Lorenz, Brown}

Heat transport has recently been proved to be very useful to probe
the spin-phonon coupling and the magnetic transitions in
multiferroic materials, including the hexagonal
$R$MnO$_3$.\cite{Sharma, Zhou, Wang_HMO, Zhao_GFO} In particular,
HoMnO$_3$ was found to have extremely strong magnetic-field
dependence of thermal conductivity ($\kappa$), which is mainly
caused by the magnetic scattering on phonons. An interesting
finding was that both the $ab$-plane and the $c$-axis magnetic
field can induce strong minimum (or ``dips") in $\kappa(H)$
isotherms, which are directly related to the transitions of
magnetic structures.\cite{Wang_HMO} In this paper, we study the
low-$T$ heat transport of TmMnO$_3$, which is a special one in
$R$MnO$_3$ family because the magnetic ions Tm$^{3+}$ do not form
a long-range order.\cite{Yen2, Skumryev, Salama} It is expected
that the magnetic correlation of Tm$^{3+}$ ions might play an
important role in the heat transport. To confirm this, the
temperature and field dependencies of $\kappa$ in TmMnO$_3$ are
compared with those in two isostructural variants YMnO$_3$ and
HoMnO$_3$. Besides rather strong magnetic-field dependencies of
$\kappa$, an important result is a sharp dip at 3.5--4 T in
$\kappa(H)$ isotherms due to the transition of Mn$^{3+}$ magnetic
structure driven by the $c$-axis field. In addition, the magnetic
field up to 14 T along the $ab$ plane does not induce any magnetic
transition. These heat transport behaviors have good
correspondence with the magnetization data of TmMnO$_3$.

\section{EXPERIMENTS}

High-quality TmMnO$_3$ single crystals were grown by using a
floating-zone technique in flowing mixture of Ar and O$_2$ with
the ratio 4:1. The crystals were carefully checked by using the
x-ray Laue photograph and cut precisely along the crystallographic
axes, with parallelepiped shape and typical size of $2.5 \times
0.6 \times 0.15$ mm$^3$ for the heat transport measurements. The
thermal conductivities were measured along both the $a$
($\kappa_{a}$) and the $c$ axes ($\kappa_c$) by using a
conventional steady-state technique with the temperature down to
0.3 K, which has been described elsewhere.\cite{Wang_HMO,
Zhao_GFO, Sun_DTN, Li_NGSO} The magnetization was measured using a
SQUID-VSM (Quantum Design). The samples for the specific-heat
measurements were cut into thin plates with the $a$ or the $c$
axis normal to the basal plane. The specific heat was measured by
the relaxation method in the temperature range from 0.4 to 30 K
using a commercial physical property measurement system (PPMS,
Quantum Design).

\section{RESULTS AND DISCUSSION}

\begin{figure}
\includegraphics[clip,width=8.5cm]{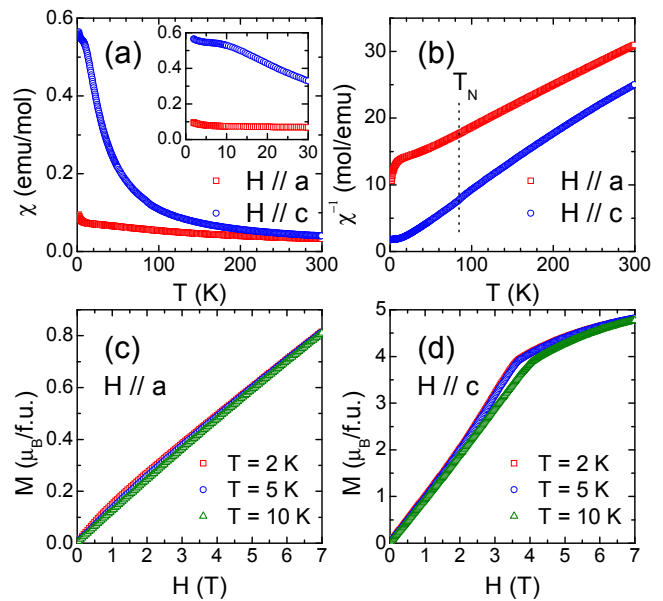}
\caption{(color online) (a, b) Temperature dependencies of the
magnetic susceptibilities and the inverse susceptibilities of
TmMnO$_3$ for magnetic field (0.1 T) applied along the $a$ and the
$c$ axes. Inset to panel (a): low-temperature data of the
susceptibilities. The dashed line in panel (b) indicated the slope
changes of data at $\sim$ 84 K, which corresponds to the N\'eel
temperature of Mn$^{3+}$ moments. (c, d) Low-temperature
magnetization curves of TmMnO$_3$ for magnetic field along the $a$
and the $c$ axes. The slope changes of $M(H)$ at 3.5--4 T in panel
(d) is related to an in-plane spin re-orientation of Mn$^{3+}$
sublattice.}
\end{figure}

Before presenting the heat transport data of TmMnO$_3$ single
crystals, we show some characterizations of the magnetization and
specific heat. Figure 1 shows the basic magnetic properties of
TmMnO$_3$ single crystals. The magnetic susceptibilities and the
inverse susceptibilities along the $a$ and the $c$ axes are
essentially the same as those in some earlier
literatures.\cite{Yen2, Skumryev} The data show two main results.
First, the temperature dependencies of the susceptibilities show
some slight changes at $\sim$ 84 K, which indicates the AF
transition of Mn$^{3+}$ moments. These changes can be more easily
detected from the temperature dependencies of the inverse
susceptibilities. Second, the Tm$^{3+}$ moments do not show
long-range order at temperatures down to 2 K. However, the low-$T$
susceptibilities show deviations from the paramagnetic behavior,
which include a kink of $\chi_c$ and a weak increase of $\chi_a$
below 10 K (see Fig. 1(b) and the inset to Fig. 1(a)). These
suggest that the magnetic correlations among Tm$^{3+}$ moments are
not negligible. The low-$T$ magnetization curves of TmMnO$_3$,
which are not available from the literature, are useful for
indicating the field-induced transitions of magnetic structure. As
shown in Fig. 1(c), the magnetic field along the $ab$ plane causes
a simple paramagnetic effect. Actually, the $M(H)$ curves are
nearly linear up to 7 T except for a small curvature at low
fields. Although there was no report on whether there is
transition of the magnetic structure for the field along the $ab$
plane, the magnetization curves shown in Fig. 1(c) strongly negate
this possibility. In contrast, the $M(H)$ in the $c$-axis field
behaves very differently. First of all, the magnetization is much
larger in the $c$-axis field, which is consistent with the
$\chi(T)$ data and indicates that the spin-easy axis of Tm$^{3+}$
ions is along the $c$ direction.\cite{Yen2, Skumryev}  In
addition, there is a clear transition at 3.5--4 T for temperatures
from 2 to 10 K. It is coincided with a field-induced transition of
Mn$^{3+}$ magnetic structure from $P6'_3c'm$ to $P6_3c'm'$,
proposed from the earlier dielectric measurements.\cite{Yen2} The
feature of $M(H)$ across this transition is the most similar to
that in ErMnO$_3$ among all hexagonal manganites.\cite{Yen2} It
needs to be pointed out that, however, there is by now no clear
explanation for the sharp change of the $M(H)$ slope at this
transition, although it should be a direct response of the
Tm$^{3+}$ moments since the Mn$^{3+}$ moments are known to have no
component along the $c$ axis. The earlier M\"{o}ssbauer spectra
indicated that the Tm$^{3+}$ ions on the 4$b$ site have AF
interactions and those on the 2$a$ site are simply
paramagnetic.\cite{Salama} It is likely that the slope changes of
magnetization curves for $H \parallel c$ is a spin flip transition
of the 4$b$ moments, considering their strong spin anisotropy. The
slower increase of magnetization above this field is the
paramagnetic contribution from the 2$a$ moments. In addition, the
magnetization for $H \parallel a$ is mainly the paramagnetic
response of the 2$a$ moments.

\begin{figure}
\includegraphics[clip,width=6.5cm]{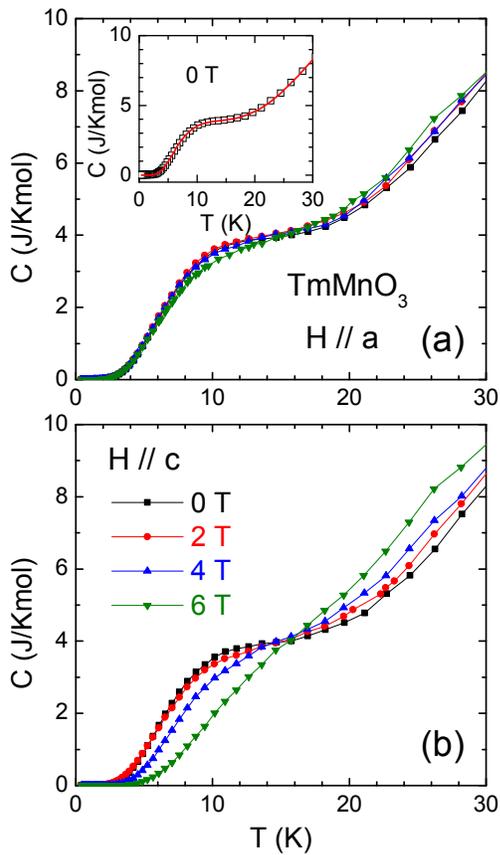}
\caption{(color online) Low-temperature specific heat of TmMnO$_3$
single crystals for magnetic fields along the $a$ and the $c$
axes, respectively. The hump at $\sim$ 4--16 K is very robust
under the in-plane field but gradually smeared out by the $c$-axis
field. Inset to panel (a): the zero-field data and a fitting
result, indicated by the solid line, by using formula (\ref{C}).}
\end{figure}

Figure 2 shows the specific-heat data of TmMnO$_3$ single crystals
at low temperatures and in magnetic fields along the $a$ and the
$c$ axes, respectively. The zero-field data indicates no phase
transition down to 0.4 K. However, there is a hump-like feature in
temperature range from 4 to about 16 K, which makes the
temperature dependence not to be described by the standard phonon
behavior. Apparently, this feature is caused by some magnetic
contributions. The magnetic fields suppress the specific heat
below $\sim$ 16 K but enhance the specific heat above $\sim$ 16 K,
signifying a Schottky anomaly from the paramagnetic ions. It is
useful to try a fitting to the zero-field data by considering both
the phononic formula and that of a simple Schottky formula for a
two-level system, that is,
\begin{equation}\label{1}
C(T) = \beta T^3 + \beta_5 T^5 + \beta_7 T^7 +
N(\frac{\Delta}{k_BT})^{2}\frac{e^{\Delta/k_BT}}{(1+e^{\Delta/k_BT})^{2}}
. \label{C}
\end{equation}
The parameters $\beta$, $\beta_5$, and $\beta_7$ describe the
phononic specific heat using the low-frequency expansion of the
Debye function $C_p = \beta T^3 + \beta_5T^5 +
\beta_7T^7$.\cite{Tari} $\Delta$ is the splitting of the
ground-state doublet of magnetic ions and the concentration of
free spins is $N/R$, with $R$ the universal gas
constant.\cite{Tari, Li_NGSO} The fitting can be pretty good, as
shown in the inset to Fig. 2(a) with the parameters $\beta = 2.6
\times 10^{-4}$ J/K$^4$mol, $\beta_5 = 1.21\times{10^{-7}}$
J/K$^6$mol, and $\beta_7 = -1.31\times {10^{-10}}$ J/K$^8$mol, $N
= 7.59$ J/Kmol, and $\Delta = 2.36$ meV. The fitting parameters
for the phononic specific heat are comparable with those in other
$R$MnO$_3$ compounds.\cite{SH_YMO, SH_RMO} The value of fitting
parameter $\Delta$ is also comparable to that from the analysis on
M\"{o}ssbauer spectra.\cite{Salama} This indicates that the
magnetic specific heat is mainly from a Schottky anomaly of
Tm$^{3+}$ ions, which do not form a long-range ordered state at
low temperatures. However, it is notable that the effect of
magnetic field on the magnetic specific heat is much larger with
$H \parallel c$ than with $H \parallel a$, showing a phenomenon
different from a pure Schottky anomaly. Actually, the field up to
6 T along the $a$ axis seems to produce very weak Zeeman splitting
of the ground-state doublet. This indicates that the short-range
magnetic correlations of Tm$^{3+}$ ions on the 4$b$ sites should
be involved. In this regard, both the specific heat and magnetic
susceptibility show a dominant paramagnetic behavior of Tm$^{3+}$
moments, but with unnegligible magnetic correlations.

\begin{figure}
\includegraphics[clip,width=6.5cm]{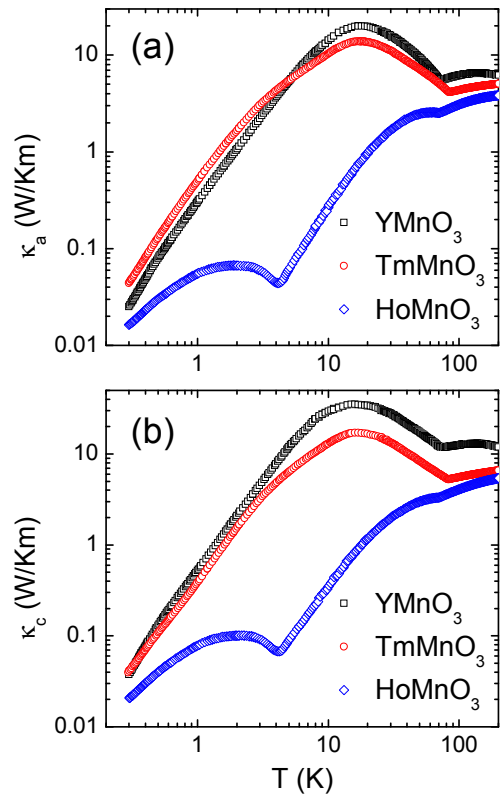}
\caption{(color online) Temperature dependencies of the $ab$-plane
and the $c$-axis thermal conductivities of TmMnO$_3$ single
crystals. The data of YMnO$_3$ and HoMnO$_3$ single crystals taken
from Ref. \onlinecite{Wang_HMO} are also shown for comparison.}
\end{figure}

Figure 3 shows the temperature dependencies of $\kappa_{ab}$ and
$\kappa_{c}$ of TmMnO$_3$ single crystals in zero field, together
with those of YMnO$_3$ and HoMnO$_3$ for
comparison.\cite{Wang_HMO} The same feature of these three
materials is that the thermal conductivities show clear ``kinks"
at the N\'eel temperatures of Mn$^{3+}$ moments, which is about 84
K for TmMnO$_3$. It is known to be due to the strong phonon
scattering by the spin fluctuations.\cite{Sharma, Zhou, Wang_HMO}
The low-$T$ heat transport behaviors of these materials seem to be
strongly affected by the magnetic properties of the rare-earth
ions. For YMnO$_3$, in which the Y$^{3+}$ ions are nonmagnetic,
the low-$T$ heat transport is a typical one in insulating
crystals,\cite{Berman} with large phonon peaks at 15 K indicating
the high quality of the crystals. For HoMnO$_3$, another strong
``dip" shows up in the $\kappa(T)$ curves at the N\'eel
temperature of Ho$^{3+}$. Due to the strong spin fluctuations
related to the multiple AF ordering and magnetic structure
transitions, the phonon heat transport of HoMnO$_3$ is so strongly
suppressed that the $\kappa$ can be 1--2 orders of magnitude
smaller than that of YMnO$_3$. For TmMnO$_3$, the Tm$^{3+}$ ions
are magnetic but there is no long-range order of their moments. As
the susceptibility and specific-heat data suggest, the magnetic
correlations or some short-range order are established among the
Tm$^{3+}$ moments at low temperatures. As a result, the magnetic
fluctuations of Tm$^{3+}$ cannot be neglected. The $\kappa(T)$
data in Fig. 3 show that although the very-low-$T$ thermal
conductivities of TmMnO$_3$ are comparable or even larger than
those of YMnO$_3$, the magnitudes of phonon peaks (also locating
at $\sim$ 15 K) are much smaller in TmMnO$_3$. In addition, the
$\kappa(T)$ curves of TmMnO$_3$ show some slight curvatures at
4--10 K, which indicates a resonant-scattering on phonons,
probably by paramagnetic ions.

\begin{figure}
\includegraphics[clip,width=5cm]{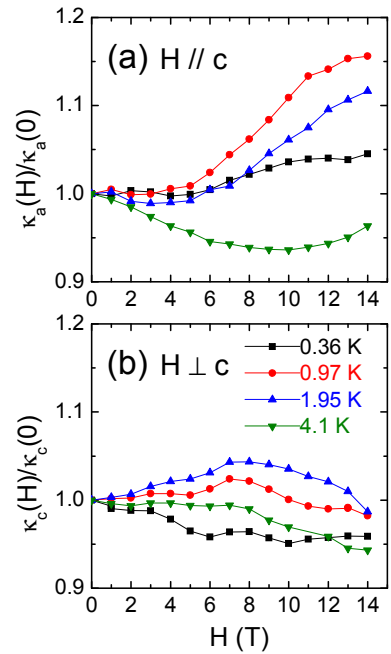}
\caption{(color online) Magnetic-field dependencies of $\kappa_a$
and $\kappa_c$ of YMnO$_3$ single crystals at low temperatures.
The magnetic field is along and perpendicular to the $c$ axis,
respectively.}
\end{figure}

Before proceeding on studying the effect of magnetic field on the
heat transport of TmMnO$_3$, we show in Fig. 4 some representative
low-$T$ $\kappa(H)$ isotherms of YMnO$_3$ single crystals, of
which the $\kappa(T)$ data were already shown in Fig. 3. It is
clear that the field dependence of $\kappa$ is not strong in
YMnO$_3$ and is probably related to some scattering effect by the
magnetic impurities or defects.\cite{Sun_PLCO, Sun_GBCO} In
passing, this magnetic scattering is likely the reason that the
very-low-$T$ $\kappa$ of YMnO$_3$ show a weaker temperature
dependence than $T^3$. Furthermore, the field dependencies of
$\kappa$ are generally rather smooth and there is no any drastic
change, which means that neither the $c$-axis nor the $ab$-plane
fields induce any transitions of the ground state and the magnetic
structure. This is in good consistency with the earlier
works.\cite{Fiebig2}

The magnetic field is found to strongly affect the heat transport
of TmMnO$_3$. As shown in Fig. 5, the effect of magnetic field on
the low-$T$ thermal conductivities depends on the direction of
field and are nearly the same for different directions of heat
current. In magnetic field along the $a$ axis, both the $\kappa_a$
and the $\kappa_c$ show an increase at subKelvin temperatures and
a decrease at higher temperatures. It is notable that the slight
curvatures of the zero-field $\kappa(T)$ curves at 4--10 K are
clearly enlarged in the 5 T and 14 T curves and a shoulder-like
feature is visualized. This confirms the phonon resonant
scattering behavior.\cite{Berman, Sun_DTN} When applying magnetic
field along the $c$ axis, the heat transport changes in a
different way. The low fields of 3.5 T and 4 T can suppress the
thermal conductivities in the whole temperature range, while a
high field of 14 T ($\parallel c$) can significantly enhance the
thermal conductivities, particularly at temperatures around 10 K.
A common feature for the $\kappa(T)$ in either $H \parallel a$ or
$H \parallel c$ is that 14 T field always increases the
conductivity at subKelvin temperatures.

\begin{figure}
\includegraphics[clip,width=8.5cm]{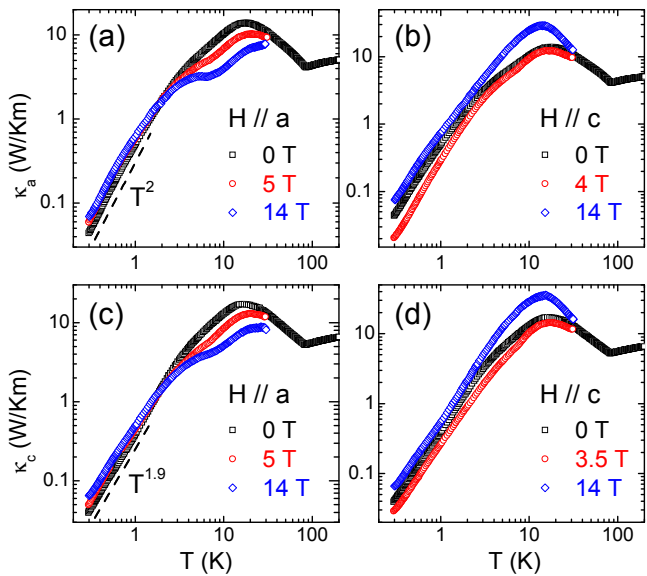}
\caption{(color online) Temperature dependencies of thermal
conductivities of TmMnO$_3$ single crystals for $\kappa_a$ (a-b)
and $\kappa_c$ (c-d) in both the zero field and several different
magnetic fields up to 14 T. The dashed lines in panels (a) and (c)
indicate the temperature dependencies of the zero-field $\kappa$
at subKelvin temperatures.}
\end{figure}

From the $\sim T^2$ dependencies of the zero-field conductivities
at subKelvin temperatures, which are much weaker than the standard
$T^3$ law of phonon conductivity at the boundary scattering limit,
it is easy to know that the microscopic phonon scatterings are
still effective at such low temperatures.\cite{Berman} Since the
crystal defects scattering on phonons usually fades away with
decreasing temperature below 1 K,\cite{Berman} the phonons are
mainly scattered by magnetic excitations or spin
fluctuations.\cite{Wang_HMO, Zhao_GFO} Therefore, the
high-field-induced increase of $\kappa$ at low temperatures
indicates that the magnetic scattering can be removed by high
magnetic field.\cite{Wang_HMO, Zhao_GFO} It can be known that the
magnetic scattering is mainly related to the Tm$^{3+}$ sublattice.
On the one hand, although the Mn$^{3+}$ moments are
antiferromagnetically ordered at low temperatures, the strong spin
anisotropy of Mn$^{3+}$ ions and the resulted anisotropy gap in
the magnon spectrum\cite{Standard, Petit, Fabreges2} make the
low-energy Mn$^{3+}$ magnon excitations impossible at very low
temperatures. On the other hand, the Tm$^{3+}$ moments do not form
a long-range ordered state but are antiferromagnetically
correlated. It is also notable that the phonon peaks of
$\kappa_a(T)$ and $\kappa_c(T)$ in 14 T field are comparable or
even larger than those of YMnO$_3$. This means that the phonon
thermal conductivities of these $R$MnO$_3$ are essentially the
same if the magnetic scattering effect is negligible.

\begin{figure}
\includegraphics[clip,width=8.5cm]{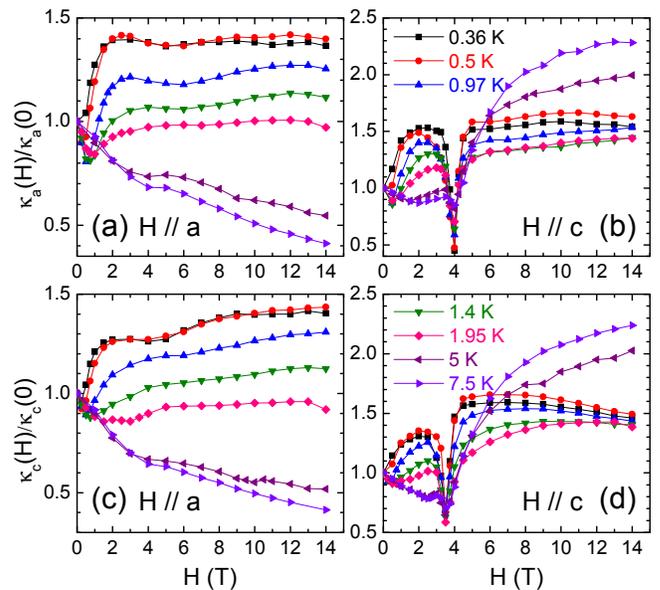}
\caption{(color online) Magnetic-field dependencies of $\kappa_a$
and $\kappa_c$ of TmMnO$_3$ single crystals at low temperatures.
The magnetic field is along either the $a$ or the $c$ axis.}
\end{figure}

The detailed magnetic-field dependencies of the low-$T$ thermal
conductivities of TmMnO$_3$ are shown in Fig. 6, which are
apparently much stronger than those of YMnO$_3$ but weaker than
those of HoMnO$_3$.\cite{Wang_HMO} First of all, as the
$\kappa(T)$ data already indicated, the field-dependencies of
$\kappa$ are mainly dependent on the direction of magnetic field
rather than that of heat current. And there are some important
features of these $\kappa(H)$ isotherms that the $\kappa(T)$ data
cannot reveal. The most remarkable behavior is the sharp ``dips"
in $\kappa(H)$ curves for $H \parallel c$ and the positions of
these minimums of $\kappa$ are temperature independent from 0.36
to 7.5 K, which are very similar to the phenomenon observed in
HoMnO$_3$.\cite{Wang_HMO} It has been known that TmMnO$_3$ has a
field-induced spin re-orientation for $H \parallel c$. The
transition field is decreased with lowering temperature from
$\sim$ 60 K and becomes weakly temperature dependent (about 4.2 T)
below 10 K.\cite{Yen2} Therefore, it is clear that the dips in
$\kappa(H)$ curves correspond to this field-induced transition of
magnetic structure. One may note that the dip fields are a bit
different in $\kappa_a(H)$ and $\kappa_c(H)$ curves, which are at
3.5 and 4 T, respectively. It is understandable if the
demagnetization effect is taken into account. The samples for
$\kappa$ measurement are shaped into long-bar-like, with the
longest dimensional along the heat current. Therefore, for
$\kappa_a$ and $\kappa_c$ samples, the $c$-axis fields are applied
to the dimensions along the thickness and length, respectively,
which yields rather different demagnetization factors for these
two measurements. According to the sample sizes, the
demagnetization factors are calculated to be 0.748 and 0.048 for
$\kappa_a$ and $\kappa_c$ samples, respectively. Using the
susceptibility data from Fig. 1, the difference in the (external)
transition fields for these two samples can be calculated and is
about 15\%,\cite{Zhao_GFO} which is in good consistent with the
experimental observation.

The $\kappa(H)$ behaves rather differently in the case of $H
\parallel a$; in particular, there is no similar sharp dip-like
feature. It is consistent with the magnetization data of Fig. 1(c)
in the regard that there is no transition of magnetic structure
for field along the $ab$ plane. The overall field dependencies of
$\kappa$, apart from the strong dips at 3.5--4 T, are also
somewhat different for $H \parallel a$ and $H \parallel c$. At the
lowest temperature of 0.36 K, the $\kappa(H)$ isotherms are very
similar in different field directions; that is, the $\kappa$ show
a quick increase at low fields and become nearly field independent
above 2 T (except for the strong dip for $H \parallel c$). With
increasing temperature up to $\sim$ 1 K, the $\kappa(H)$ isotherms
in different field directions are still rather similar, including
the appearance of a small dip at low fields. Actually the features
of a low-field dip and a high-field plateau in $\kappa(H)$ curves
are qualitatively the same as that caused by the paramagnetic
scattering on phonons.\cite{Sun_PLCO, Sun_GBCO, Li_NGSO} A
significant difference between $H \parallel a$ and $H \parallel c$
appears at high temperatures. It can be seen that $\kappa(H)$ at 5
and 7.5 K show continuous suppressions for $H \parallel a $ and
strong high-field increases for $H \parallel c$, respectively. It
seems that this difference has some correspondence with the field
dependencies of specific heat shown in Fig. 2. Therefore, it is
possible to understand this result using the paramagnetic
scattering scenario,\cite{Sun_PLCO, Sun_GBCO, Li_NGSO} in which
the magnetic scattering can be strong in the former case but would
be removed in the latter case.

At last, it would be useful to compare the heat transport of
TmMnO$_3$ with that of HoMnO$_3$. In an earlier work, some
dip-like anomalies have also been observed in the low-$T$
$\kappa(H)$ isotherms of HoMnO$_3$ at the field-induced
transitions of magnetic structure.\cite{Wang_HMO} Those results
suggested that Ho$^{3+}$ moments form two AF sublattices and
undergo two transitions upon increasing field along the $c$ axis,
accompanied with two spin re-orientations of the Mn$^{3+}$
sublattice. In addition, the $\kappa(H)$ data for $H
\parallel ab$ revealed that a spin re-orientation of Mn$^{3+}$
sublattice could also occur even when the Ho$^{3+}$ sublattice may
not be affected. In contrast, the magnetic transitions evidenced
by the heat transport as well as the magnetization are simpler in
TmMnO$_3$. Another difference between TmMnO$_3$ and HoMnO$_3$ is
that the magnetic-field dependencies of $\kappa$ are at least
several times larger in HoMnO$_3$. In fact, HoMnO$_3$ has very
small zero-field $\kappa$ and the significant increase of
conductivity at high magnetic field.\cite{Wang_HMO} Apparently,
the mechanism for the magnetothermal conductivity can be shared
for these two compounds, while the difference of $\kappa(H)$ is
due to the stronger $R^{3+}$-Mn$^{3+}$ magnetic interactions in
HoMnO$_3$. In this regard, thermal conductivity results provide a
transparent demonstration that the spin-phonon coupling in
$R$MnO$_3$ is mainly determined by the magnetism of $R^{3+}$ and
is the strongest in HoMnO$_3$.

Although the mechanism of the field-induced re-orientation of
Mn$^{3+}$ moments is still not very clear, the magnetic
interaction between rare-earth ions and Mn$^{3+}$ ions is
obviously very important. In hexagonal $R$MnO$_3$, the exchange
coupling between $R^{3+}$ and Mn$^{3+}$ ions is weak because their
moments are perpendicular to each other. In such case, the dipolar
interaction is more important and larger magnetic moment of
Ho$^{3+}$ results in stronger interaction.\cite{Yen2}
Nevertheless, the magnetic interactions can be described by the
effective fields acting on $R^{3+}$ and Mn$^{3+}$ moments. For
magnetic field along the $c$ axis, there is no doubt that the
transitions of $R^{3+}$ sublattices cause the re-orientations of
Mn$^{3+}$ moments, because the Mn$^{3+}$ moments are so strongly
confined in the $ab$ plane that they cannot be in-plane rotated by
the magnetic field. It is likely that when the $R^{3+}$ moments
are re-aligned in the magnetic field, the effective field from the
Mn$^{3+}$ lattice may cause an increase of the $R^{3+}$ energy.
The choice of a different configuration of the Mn$^{3+}$ spin
lattice would be an effective way to lower the magnetic energy and
stabilize the magnetic structure. The case for magnetic field
along the $ab$ plane could be more complicated. The
re-orientations of the Mn$^{3+}$ moments would be irrelevant to
the $R^{3+}$ ions if they have extremely strong Ising anisotropy.
However, the different magnetic transitions between TmMnO$_3$ amd
HoMnO$_3$ clearly indicate that the rare-earth ions are playing an
important role also for $H \parallel ab$. One possibility is that
the $R^{3+}$ moments can be canted or titled to the $ab$ direction
by the magnetic field, which results in a change of the effective
field on the Mn$^{3+}$ moments and the instability of their
orientation. Further quantitative theoretical investigations on
the $R^{3+}$-Mn$^{3+}$ magnetic interactions are needed.

\section{SUMMARY}

The low-temperature heat transport of the hexagonal TmMnO$_3$
shows strong magnetic-field dependencies, which points to a rather
strong spin-phonon coupling in this material. A sharp dip-like
feature in $\kappa(H)$ isotherms for $H \parallel c$ is found to
be due to the transition of magnetic structure, as the
magnetization shows. The absence of the sharp transition of
$\kappa$ for $H \parallel a$ indicates that the in-plane field up
to 14 T cannot change the magnetic structure. The comparison of
transport properties among TmMnO$_3$, YMnO$_3$ and HoMnO$_3$
confirms that the magnetisms of rare-earth ions are the key to
determine the $H-T$ phase diagram and the strength of spin-phonon
coupling.

\begin{acknowledgements}

This work was supported by the National Natural Science Foundation
of China, the National Basic Research Program of China (Grant Nos.
2009CB929502 and 2011CBA00111), and the Fundamental Research Funds
for the Central Universities (Program No. WK2340000035).

\end{acknowledgements}

\end{document}